\documentclass[12pt,preprint]{aastex}
\usepackage{url}

\newcommand{\fig}[1]{Figure~\ref{#1}}

\newcommand{\speed}[1]{#1 km~s${}^{-1}$}

\newcommand{\acc}[1]{#1 m~s${}^{-2}$}

\begin{document}

\shorttitle{} %

\shortauthors{Shen et al.}

\title{Simultaneous Observations of a Large-Scale Wave Event in the Solar Atmosphere: From Photosphere to Corona}

\author{Yuandeng Shen\altaffilmark{1,2}, and Yu Liu\altaffilmark{1}}

\altaffiltext{1}{Yunnan Astronomical Observatory, Chinese Academy of Sciences, Kunming 650011, China; ydshen@ynao.ac.cn}
\altaffiltext{2}{Graduate University of Chinese Academy of Sciences, Beijing 100049, China}

\begin{abstract}
For the first time, we report a large-scale wave that was observed simultaneously in the photosphere, chromosphere, transition region and low corona layers of the solar atmosphere. Using the high temporal and high spatial resolution observations taken by the Solar Magnetic Activity Research Telescope at Hida Observatory and the Atmospheric Imaging Assembly (AIA) onboard {\em Solar Dynamic Observatory}, we find that the wave evolved synchronously at different heights of the solar atmosphere, and it propagated at a speed of \speed{605} and showed a significant deceleration (\acc{-424}) in the extreme-ultraviolet (EUV) observations. During the initial stage, the wave speed in the EUV observations was \speed{1000}, similar to those measured from the AIA 1700 \AA\ (\speed{967}) and 1600 \AA\ (\speed{893}) observations. The wave was reflected by a remote region with open fields, and a slower wave-like feature at a speed of \speed{220} was also identified following the primary fast wave. In addition, a type-II radio burst was observed to be associated with the wave. We conclude that this wave should be a fast magnetosonic shock wave, which was firstly driven by the associated coronal mass ejection and then propagated freely in the corona. As the shock wave propagated, its legs swept the solar surface and thereby resulted in the wave signatures observed in the lower layers of the solar atmosphere. The slower wave-like structure following the primary wave was probably caused by the reconfiguration of the low coronal magnetic fields, as predicted in the field-line stretching model.
\end{abstract}


\section{INTRODUCTION}
Large-scale wave-like perturbations in the solar atmosphere have been observed for many years. For example, the H$\alpha$ Moreton wave \citep{more60,bala07,bala10,gilb08,naru08,muhr10}, the \ion{He}{2} 10830 \AA\ wave \citep{vrsn02,gilb04a,gilb04b}, the extreme-ultraviolet (EUV) wave \citep{thom98,long08,gopa09,pats09,liu10,vero10,vero11} and the soft X-ray (SXR) wave \citep{khan02,naru02,naru04,huds03,warm05}. The H$\alpha$ Moreton wave, which manifests as a propagating dark/white front in the H$\alpha$ off-band Dopplergrams, has been recognized as a chromospheric surface wave observed immediately following an impulsive flare \citep{more60,atha61,uchi68,naru04}. Observations have indicated that the \ion{He}{2} 10830 \AA\ and SXR waves are consistent with the chromospheric Moreton wave and thereby they were interpreted as fast-mode waves and were thought to be the counterparts of the Moreton wave at different heights \citep{gilb04b,naru02,naru04,warm05}. For the EUV waves, significant controversy remains over their physical natures and origins. So far, there are several competing interpretations for the EUV waves, including the fast-mode wave model \citep{wang00,wu01,warm01,ofma02,schm10}, the slow-mode wave model \citep{will07}, and the non-wave models which are related to a current shell or successive restructuring of field lines caused by coronal mass ejections (CMEs) \citep{dela00,dela07,dela08,chen02,chen05,chen11,attr07,attr10}. In addition, a few authors proposed that both the wave and non-wave models should be required to explain the complex EUV waves \citep[e.g.,][]{zhuk04,cohe09,liu10,down11}. Detailed observational characteristics and various theoretical explanations of the EUV waves could be found in several recent reviews \cite[e.g.,][]{warm10,gall11}.

At present both the wave and non-wave models could not fully explain the observed characteristics of the EUV waves. Therefore, investigating the temporal and spatial evolutions of the EUV waves using high resolution, multiwavelength observations should be an effective way to clarify their physical natures and origins. In this letter, we present the observations of an EUV wave on 2011 August 9. Although this event has been reported by \cite{asai12}, they only studied the wave in H$\alpha$ and 193 \AA\ observations. In this letter, for the first time we report the wave signatures observed simultaneously in the photosphere, chromosphere, transition region as well as the corona, using the ultraviolet (UV) and EUV observations taken by the Atmospheric Imaging Assembly \citep[AIA;][]{leme12} onboard the {\em Solar Dynamics Observatory}. We find that the wave signatures at different height of the solar atmosphere were caused by a fast magnetosonic shock wave propagating in the corona.

\section{INSTRUMENTS AND DATA SETS}
The full-disk H$\alpha$ images are obtained by the Solar Magnetic Activity Research Telescope \citep[SMART;][]{ueno04} at Hida Observatory, Kyoto University, Japan. The SMART provides full-disk H$\alpha$ images in seven channels: H$\alpha$ center and six off-bands ($\pm 0.5$, $\pm 0.8$, and $\pm 1.2$ \AA). Its cadence is 2 minutes, and the pixel size is $0\arcsec.56$. The AIA onboard the {\em SDO} has high time resolution of up to 12 (24) s for the EUV (UV) channels, and the images have a pixel resolution of $0\arcsec.6$. In this letter, all the AIA's UV (1700 and 1600 \AA) and EUV (94, 131, 171, 193, 211, 304, and 335 \AA) observations are used to analysis the wave kinematics at different heights of the solar atmosphere. All images are differentially rotated to a reference time (08:05:00 UT), and the solar north is up, west to the right.

\section{RESULTS}
The wave event on 2011 August 9 was accompanied by a {\em GOES} X6.9 flare in NOAA AR11263 (N18W80) and a halo CME with a average speed (acceleration) of \speed{1610} (\acc{-40})\footnote{\url{http://cdaw.gsfc.nasa.gov/CME_list}}. The flare started at 07:48 UT and peaked at 08:05 UT, which is the most powerful flare observed so far in the current solar cycle 24. In this letter, we mainly investigate the kinematics and the spatial correlation of the wave at different heights of the solar atmosphere.

\begin{figure}\epsscale{0.8}
\plotone{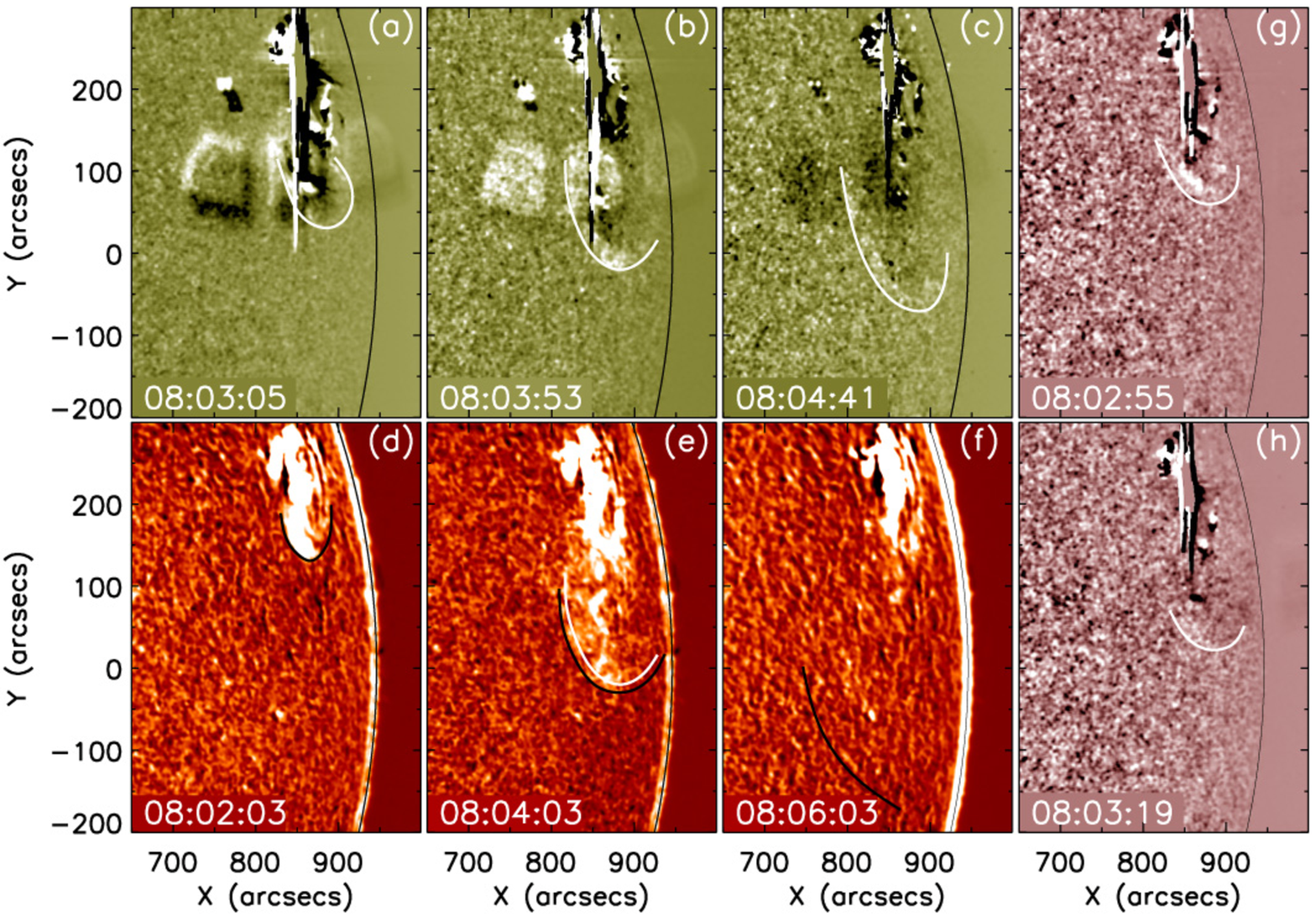}
\caption{AIA 1600 \AA\ (a -- c), 1700 \AA\ (g -- h), and SMART H$\alpha$ center (d -- f) base-difference images show the wave signatures in the transition region, photosphere, and chromosphere of the solar atmosphere respectively. The thick curves outline the wavefront, while the thin curves marks the disk limb (the same in \fig{fig2}). The white contour in panel (e) is the wavefront determined from the 1600 \AA\ image at 08:03:53 UT. The field of view (FOV) for each frame is $350\arcsec \times 500\arcsec$. Animations (Animation 1 -- 3) for this figure are available in the online version of the journal.  \label{fig1}}
\end{figure}

\fig{fig1} shows the morphological evolution of the wave on the 1600 and 1700 \AA, and H$\alpha$ center base-difference images, in which the thick curves outline the wavefronts at different times. The wavefront determined from the 1600 \AA\ image at 08:03:53 UT is overlaid on the H$\alpha$ image at 08:04:03 UT (white curve in \fig{fig1}(e)). One can see that the wavefronts at the two moments showed a similar shape, which indicates the synchronous evolution of the wave in different atmosphere layers. The wavefront seeing on the 1700 \AA\ images was relatively weak. The end times of the wave in the 1700 \AA, H$\alpha$, and 1600 \AA\ observations were at 08:04:55, 08:10:03, and 08:06:41 UT, respectively. Assuming the wave started at 08:02:00 UT, the corresponding lifetimes of the wave at different lines should be 175 (1700 \AA), 483 (H$\alpha$), and 281 s (1600 \AA) (see Animations 1 -- 3).

\begin{figure}\epsscale{0.8}
\plotone{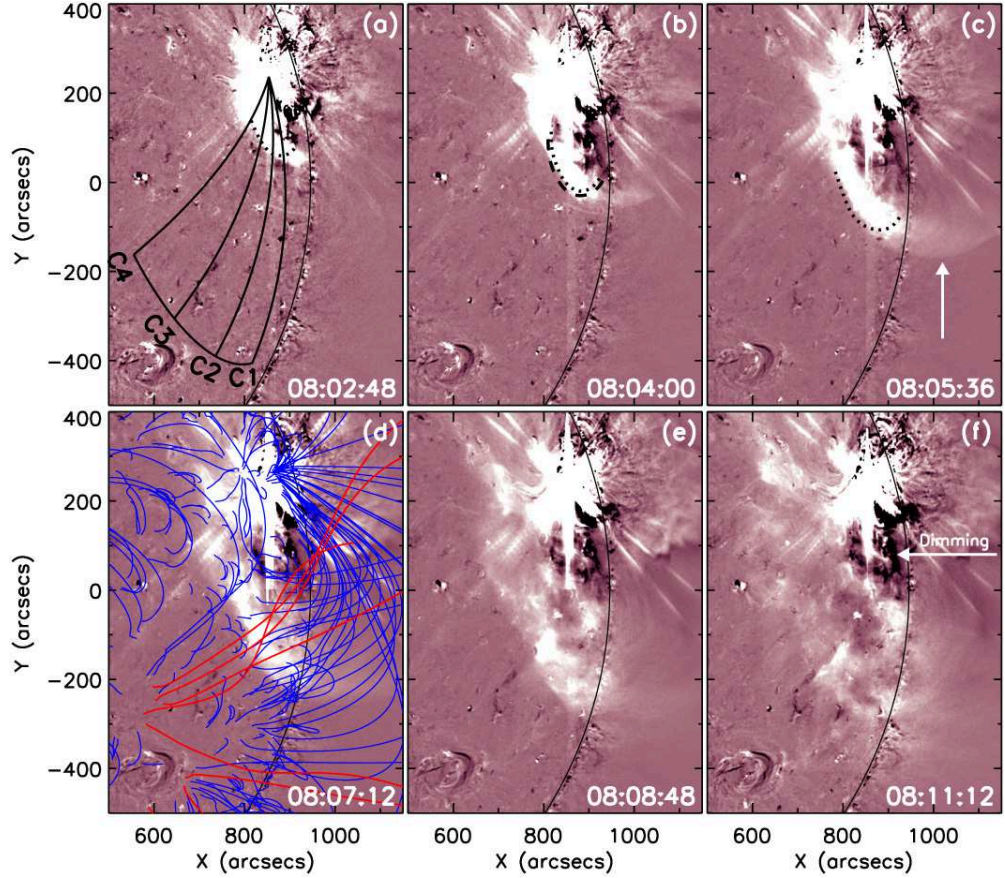}
\caption{Time sequence of 211 \AA\ base-difference images show the morphologic evolution of the wave in the corona. Black curves C1 -- C4 are great circles of the solar surface passing through the flare kernel, which are used to obtain the time-distance diagrams as shown in Figures 3 and 4. The dotted curve in panel (a) is the wavefront determined from the 1700 \AA\ image at 08:02:31 UT; the dotted curves in panels (b) and (c) are the wavefronts obtained from the 1600 \AA\ images at 08:03:53 UT and 08:05:29 UT, respectively; the dashed curve in panel (b) is the wavefront determined from the H$\alpha$ image at 08:04:03 UT. The white arrow in panel (c) points to the expanding dome, while the white arrow in panel (f) points to the dimming region. The extrapolated potential magnetic field lines are overlaid in panel (d), in which the blue and red lines represent the closed and open field lines respectively. The FOV for each frame is $650\arcsec \times 900\arcsec$. An animation (Animation 4) is available in the online version of the journal.
\label{fig2}}
\end{figure}

The wave in EUV observations is shown in \fig{fig2}, using the 211 \AA\ base-difference images. It was first observed as a semicircular sharp emission at 08:02:48 UT, and the position of the wavefront coincided well with those observed at H$\alpha$ and UV lines (see the overlaid dotted and dashed curves in \fig{fig2}(a) -- (c) and Animation 4). Moreover, a dome-like structure, which extended from the sharp bright wavefront, can be observed off the disk limb (see the white arrow in \fig{fig2}(c)), which is thought to be a shock wave traveling in the corona \citep[also see,][]{asai12}. Therefore, the sharp bright wave could be considered as the intersection of the shock wave with the corona. After 08:07:12 UT, the sharp bright wavefront became more and more diffuse (see \fig{fig2}(d) -- (e)). It is interesting that a dimming region was observed behind the bright wavefront, which did not expand to a large distance as the bright wavefront did. This result supports the scenario that the dimming region maps the CME footprint on the solar surface, while the EUV wave is a shock wave driven by the associated CME \citep{pats09,muhr10,temm11}. To reveal the magnetic topology of the coronal condition where the wave propagated, we extrapolate the three-dimensional coronal magnetic field using the potential field source surface \citep[PFSS;][]{schr03} model based on the Helioseismic and Magnetic Imager \citep[HMI;][]{scho12} magnetograms. The extrapolated field lines are overlaid in \fig{fig2}(d), from which we find that a remote region with open fields was situated on the propagation path of the wave (see the red lines in \fig{fig2}(d)).

\begin{figure}\epsscale{0.8}
\plotone{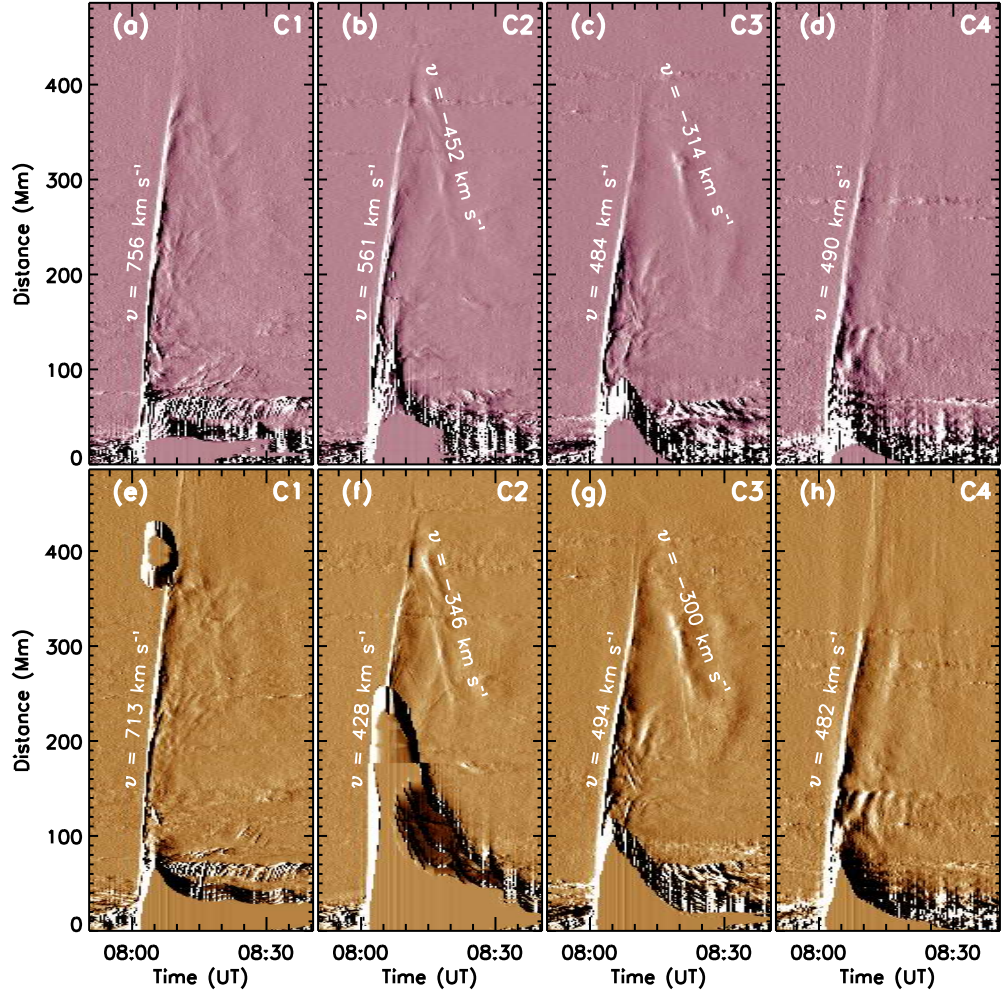}
\caption{Time-distance diagrams along cuts C1 -- C4 obtained from the 211 \AA\ (a -- d) 193 \AA\ (e -- h) running difference images. \label{fig3}}
\end{figure}

The measurements of the wave kinematics along cuts C1 -- C4 are shown in \fig{fig3}, using the time-distance diagrams obtained from the 211 and 193 \AA\ running difference images. In each time-distance diagram, the propagating wave could be identified as a bright stripe of positive slope. A linear fit to it yields the wave speed in the plane of the sky. Obviously, the wave propagated with different speeds along different cuts (428 -- \speed{756}, see \fig{fig3}). It is interesting that another wave stripe of negative slope was observed in the time-distance diagrams obtained from cuts C2 and C3. As it can be seen in \fig{fig2}(d), cuts C2 and C3 are passing through the region with open magnetic fields. This suggests that the stripe of negative slope represents the reflected wave from the open fields region. The speed of the reflected wave was 300 -- \speed{452}, which indicates that the primary wave has been decelerated significantly when it reached the open fields region. More importantly, the reflection effect manifested the wave nature of the primary EUV wave.

\begin{figure}\epsscale{0.8}
\plotone{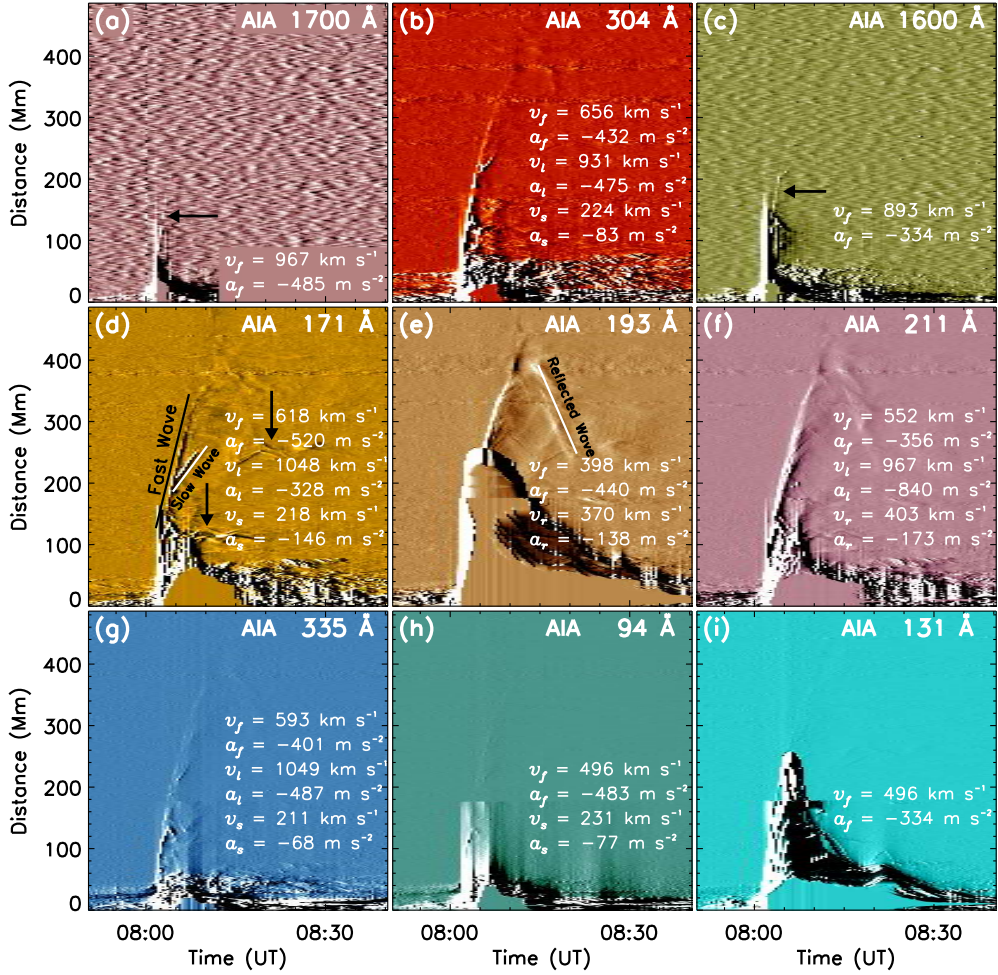}
\caption{Time-distance diagrams obtained from the running difference images show the wave kinematics along cut C2 in all AIA UV and EUV channels. In this figure, $v_{\rm f}$ ($a_{\rm f}$) is the speed (acceleration) of the primary fast wave; $v_{\rm l}$ ($a_{\rm l}$) is the speed (acceleration) of the primary fast wave during the initial stage (100 $\leqslant d \leqslant$ 200 Mm); $v_{\rm s}$ ($a_{\rm s}$) is the speed (acceleration) of the slower wave; and $v_{\rm r}$ ($a_{\rm r}$) is the speed (acceleration) of the reflected wave. A few oscillating structures are indicated by the black arrows in panel (d). \label{fig4}}
\end{figure}

To compare the wave kinematics at different heights of the solar atmosphere, the time-distance diagrams (along cut C2) obtained from the running difference images of all the UV and EUV channels are shown in \fig{fig4}. In the 1700 and 1600 \AA\ time-distance diagrams, the propagating wave is identified as a faint stripe that could only be traced to about 200 Mm from the flare kernel (see the black arrows in \fig{fig4}(a) and (c)). The speeds (accelerations) of the wave measured from the 1700 and 1600 \AA\ diagrams are \speed{967} (\acc{-485}) and \speed{893} (\acc{-334}), respectively. In the time-distance diagrams obtained from the EUV observations, the wave can be traced to about 400 Mm from the flare kernel, and the wave speed was 398 -- \speed{656}, which averaged at \speed{605}. Due to the influence of the flare diffraction, we do not take the the speeds measured from 193, 94, and 131 \AA\ observations into account. To compare the wave speed in the EUV and UV observations, we calculate the wave speed in the EUV observations within a distance of 100 -- 200 Mm from the flare kernel, and find that the wave propagated at a speed of 967 -- \speed{1049} (averaged at \speed{999}) within this distance, similar to the wave speed measured from the 1600 and 1700 \AA\ observations. This result indicates that the wave signature observed at different heights of the solar atmosphere resulted from the same physical origin and evolved synchronously. The acceleration of the wave measured from the EUV observations was -334 -- \acc{-520} (averaged at \acc{-424}). Meanwhile, the average acceleration of the wave during the initial stage was \acc{-533}, larger than that measured within the whole lifetime of the wave, which suggests the rapidly deceleration of the wave during its initial stage. It is important to note that a slower wave is observed in the EUV time-distance diagrams (see \fig{fig4}(b), (d), (g) and (h)). It propagated at a speed of 211 -- \speed{231} and with a small acceleration of -68 -- \acc{-146}. The average speed of the slower wave was \speed{220}, about three times smaller than that of the primary fast wave. We think that this slower wave was probably caused by the reconfiguration of the low coronal fields due to the eruption of the associated CME, as it has been predicted in the field-line stretching model \citep{chen02,chen05}.

\begin{figure}\epsscale{0.8}
\plotone{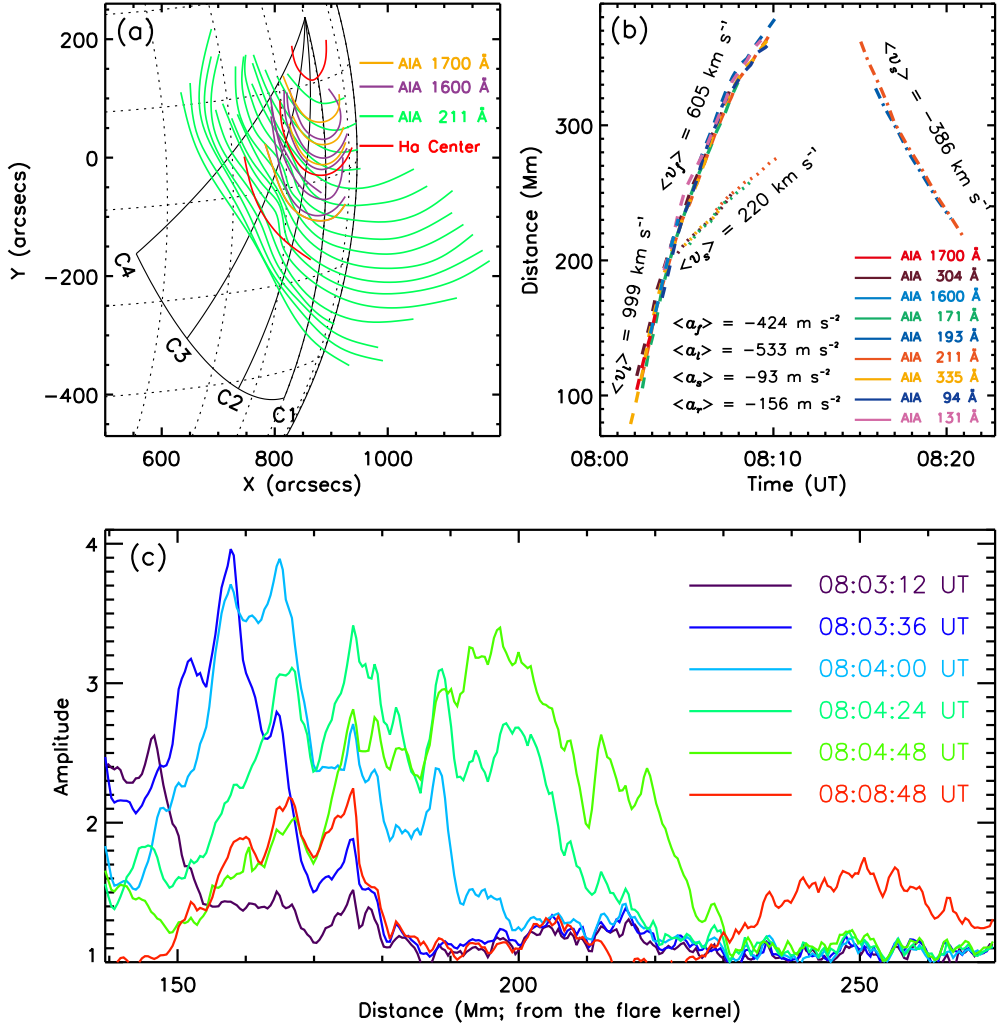}
\caption{Panel (a) shows the propagating wavefront at different times, in which the yellow, purple, green, and red curves represent the wavefront determined from the 1700, 1600, and 211 \AA, and H$\alpha$ center observations, respectively. Panel (b) shows the wave stripes at different wavelengths, which are determined from \fig{fig4}, and the average wave speed and acceleration along cut C2 are also plotted. Panel (d) is a plot of the perturbation profiles along cut C2 within a distance of 140 -- 270 Mm from the flare kernel. Note that the perturbation profiles are obtained from ratio images, where each frame is divided by a pre-event frame. The corresponding times of the perturbation profiles are indicated by the different colors.
\label{fig5}}
\end{figure}

All the wavefronts determined from the 1700 (yellow), 1600 (purple), and 211 \AA\ (green) and H$\alpha$ (red) observations are plotted in \fig{fig5}(a). One can see that all of them showed a semicircular shape and propagated synchronously on the solar surface, and the wavefronts determined on 1700, 1600 \AA, and H$\alpha$ images coincided well with the sharp bright wavefront observed in the 211 \AA\ observations. On the other hand, the wave stripes in the time-distance diagrams along C2 at different wavelengths are all plotted in \fig{fig5}(b). It can be seen that the primary fast wave has a similar speed along cut C2 in all UV and EUV channels, which suggests a common mechanism of the wave in the different layers of the solar atmosphere.

The perturbation profiles of the wavefront at different times are plotted in \fig{fig5}(c), which well reflected the evolution of the wave. It can be seen that the steepness of the perturbation profile weakened quickly with increasing time, while the amplitude first showed an increase and then decreased significantly within several minutes. The amplitude reached the highest value of 3.96 at 08:03:36 UT and dropped to 1.75 at 08:08:48 UT. Assuming the intensity enhancement is primarily due to plasma compression and the wave propagation was perpendicular to the magnetic fields, we can obtain the magnetosonic Mach number of the wave \citep{prie82}, which is 1.85 (1.25) at 08:03:36(08:08:48) UT. On the other hand, the width of the perturbation profile of the wavefront broadened a lot with increasing time. All these results including the deceleration of the magnetosonic Mach number, amplitude decrease, and wavefront broadening are consistent with the shock wave scenario \citep{prie82,vero10}. In addition, the evidence for the appearance of the shock wave was also observed as a type-II radio burst in the metric radio spectrogram, which was observed from 08:02:40 to 08:06:30 UT and had a derived speed of \speed{850} \citep[see;][for details]{asai12}. The start time and the speed of the type-II radio burst were in agreement with those observed in the UV and EUV observations.

\section{CONCLUSIONS AND DISCUSSIONS}
For the first time, we report an EUV wave that was observed simultaneously in the different layers of the solar atmosphere. With the high temporal and spatial resolution H$\alpha$, UV and EUV observations taken by the SMART and the {\em SDO}/AIA, we investigate the kinematics and the spatial correlation of the wave at different heights of the solar atmosphere. The main results are summarized as follows.
\begin{enumerate}
  \item The wave could be observed simultaneously in the photosphere, chromosphere, transition region and the low corona. The lifetime of the wave determined from the 1700 \AA, 1600 \AA, and H$\alpha$ observations were 175, 483, and 281 s, respectively. The wave signatures observed in the photosphere, chromosphere, transition region and the sharp bright wavefront observed in the EUV observations evolved synchronously in space and time, which suggests that these waves signature at different layers of the solar atmosphere had a common origin and the same physical mechanism.
  \item In the EUV observations, the average speed (acceleration) of the wave was \speed{605} (\acc{-424}). During the initial stage, the wave had a similar speed of about \speed{1000} at all UV, EUV, H$\alpha$, and radio wavelength bands. Moreover, the wave speed during the initial stage decreased faster than that during the whole wave lifetime.
  \item The wave kept propagating after the following dimming region has stopped to expand. This phenomenon suggests that the wave was driven by the associated CME \citep{pats09,muhr10,temm11} and thereby rules out the pseudo-wave models \citep[e.g.,][]{dela07,dela08,attr07}. In addition, the wave nature of the EUV wave was also manifested by the reflection of the EUV wave from the remote open fields region and the type-II radio burst observed in the metric radio spectrogram.
  \item A slower wave with a speed (acceleration) of \speed{220} (\acc{-93}) was observed behind the primary fast wave, which was probably resulted from the reconfiguration of the low coronal fields caused by the associated CME \citep{chen02,chen05}.
  \item During the initial stage, the steepness of the wavefront's perturbation profile weakened quickly and the width broadened significantly, and the amplitude first increased a lot and then decreased significantly within several minutes. The highest Mach number of the wave is 1.85, which quickly decreased to a Mach number of unity. All these results are consistent with the physical properties of a fast magnetosonic shock wave \citep{prie82,warm01}. In addition, the evidence for appearance of the shock wave was also identified, such as the dome-like structure observed in the EUV observations and the type-II radio burst in the metric radio spectrogram.
\end{enumerate}

Base on our analysis results, we conclude that the EUV wave analyzed in this letter should be a fast magnetosonic shock wave, which was firstly driven by the associated CME and then propagated freely in the corona. The wave signatures observed in the photosphere, chromosphere, transition region layers and the sharp bright wavefront observed in the corona were the intersections of the coronal shock wave with these lower layers of the atmosphere. As the shock propagates, its energy is gradually dissipated into the ambient plasma medium by either plasma oscillations (that are subsequently damped) or plasma microinstabilities \citep{prie82}. This energy dissipation and the expansion of the shock front will cause the decrease the perturbation amplitude, and consequently its speed and Mach number. This also explains why the deceleration rate decreases with increasing time and distance. Eventually, the shock will decay to an ordinary fast-mode wave with Mach number of 1 \citep{warm04}.

In the field-line stretching model \cite{chen02,chen05}, as the CME flux rope rises, a piston-driven shock wave is formed preceding the envelope of the expanding CME, which sweeps the solar surface at a super-Alfv\'{e}nic speed. This mechanism implies that the legs of the shock wave would produce the surface wave phenomena in the solar lower atmosphere layers. In the meantime, a slower wave-like feature at a speed of about three times smaller than that of the shock wave could be identified behind the fast shock wave. This structure is thought to be produced by the successive stretching of closed field lines during the launch of the CME. This model also predicts the generation of type-II radio bursts from the top of the shock wave. Our observational results are not only consistent with this field-line stretching model, but also indicate that the legs of the shock wave can extend downward into the photosphere. Further investigations involving high temporal and spatial resolution observations of UV, EUV, and SXR would be helpful to fully understand the physical natures of the waves in the solar atmosphere.

\acknowledgments {\em SDO} is a mission for NASA's Living With a Star (LWS) Program. We thank the SMART for data support. The authors thank an anonymous referee for many constructive suggestions. This work is supported by the Natural Science Foundation of China under grants 10933003, 11078004, and 11073050, and the National Key Research Science Foundation (2011CB811400).


\end{document}